\documentclass[10pt,conference]{IEEEtran}
\IEEEoverridecommandlockouts
\usepackage{cite}
\usepackage{amsmath,amssymb,amsfonts}
\usepackage{graphicx}
\usepackage{textcomp}
\usepackage{xcolor}

\usepackage{listings}
\usepackage{capt-of}

\usepackage{booktabs} %
\usepackage{wrapfig}
\usepackage{algorithm}
\usepackage{algpseudocode}
\usepackage{multirow}
\usepackage{balance}
\usepackage{color}
\usepackage{xspace}

\usepackage{textpos}

\newcommand{\MT}{MT\xspace}

\newcommand{\CR}[1]{\textcolor{black}{#1}}

\newcounter{cLINE}

\newcommand{\TEXTTT}[1]{$\mathtt{#1}$}
\newcommand{\MR}{MR\xspace}
\newcommand{\MRs}{MRs\xspace}

\newcommand{\SMRL}{SMRL\xspace}

\newcommand{\CHANGED}[1]{\textcolor{black}{#1}}
\newcommand{\CHANGEDLAST}[1]{\textcolor{black}{#1}}

\newcommand{\manuallabel}[2]{\def\@currentlabel{#2}\label{#1}}

\usepackage{amsmath}
\usepackage{tikz}

\def\BibTeX{{\rm B\kern-.05em{\sc i\kern-.025em b}\kern-.08em
    T\kern-.1667em\lower.7ex\hbox{E}\kern-.125emX}}

\begin{document}

\title{Metamorphic Security Testing for Web Systems}

\author{\IEEEauthorblockN{Phu X. Mai\IEEEauthorrefmark{1}, Fabrizio Pastore\IEEEauthorrefmark{1}, Arda Goknil\IEEEauthorrefmark{1}, Lionel Briand\IEEEauthorrefmark{1}\IEEEauthorrefmark{2}}
\IEEEauthorblockA{\IEEEauthorrefmark{1}SnT Centre for Security, Reliability and Trust, University of Luxembourg, Luxembourg}
\IEEEauthorblockA{\IEEEauthorrefmark{2}School of Engineering and Computer Science, University of Ottawa, Canada}
\{xuanphu.mai,fabrizio.pastore,arda.goknil,lionel.briand\}@uni.lu
}

\maketitle

\begin{abstract}

\CR{Security testing 
verifies that the data and the resources of software systems are protected from attackers. Unfortunately, it suffers from the oracle problem, which refers to the challenge, given an input for a system, 
of distinguishing correct from incorrect behavior.}
In many situations where potential vulnerabilities are tested, a test oracle may not exist, or it 
might be impractical due to the many inputs for which specific oracles have to be defined.

\CR{In this paper, we propose a metamorphic testing approach that alleviates the oracle problem in security testing. It enables engineers to specify 
metamorphic relations (MRs) that capture security properties of the system. 
Such MRs are then used to automate testing and detect vulnerabilities.}

We provide a catalog of 22 system-agnostic MRs to automate security testing in Web systems. 
\CR{Our approach targets 39\% of the OWASP security testing activities not automated by state-of-the-art techniques.
It automatically detected 10 out of 12 vulnerabilities affecting two widely used systems, one commercial and the other open source (Jenkins).}

\end{abstract}

\begin{IEEEkeywords}
Software Engineering, Software Security
\end{IEEEkeywords}

\begin{textblock*}{181.5mm}(.0\textwidth,-12.5cm)
\centering{\textbf{To be published in the Proceedings of the IEEE International Conference on Software Testing, Verification and Validation (ICST) 2020}}
\end{textblock*}

\begin{textblock*}{181.5mm}(.0\textwidth,12.5cm)
\textbf{\textsuperscript{\textcopyright} \textsuperscript{\textcopyright} 2020 IEEE. Personal use of this material is permitted. Permission from IEEE must be obtained for all other uses, in any current or future media, including reprinting/republishing this material for advertising or promotional purposes, creating new collective works, for resale or redistribution to servers or lists, or reuse of any copyrighted component of this work in other works.}
\end{textblock*}

\section{Introduction}

Security testing aims to uncover flaws in software mechanisms that protect data and ensure the delivery of the intended system functionality. 
It is driven by security requirements which encompass both security properties of the system and the prevention of potential security threats~\cite{Haley2008,Mouratidis2007,Felderer2016,Mai2018a,ISSRE}. 
In contexts where test case execution is automated, an automated \textit{test oracle} (i.e., a mechanism for determining whether a test case has passed or failed) is needed to check the execution result. It often consists of comparing expected and observed outputs. 

Security test cases seldom rely on automated test oracles, most often because it is infeasible or impractical to specify them due to a large number of test inputs. 
\CHANGED{In other words, \emph{security testing suffers from the oracle problem}~\cite{Barr2015,Staats2011,Pezze2014}, which refers to situations where it is extremely difficult or impractical to determine the correct output for a given test input.}  
For instance, a security test case for the bypass authorization schema vulnerability should verify, for every specific user role, whether it is possible to access resources that should be available only to a user who holds a different role~\cite{OWASPtesting}.
This type of vulnerability can often be discovered by verifying 
\CHANGED{the access to various resources with different privileges and roles.}
However, questions arise when defining oracles. What are the resources that can only be accessible by a user with a specific role or privilege? Are the test outputs consistent with expectations regarding accessibility?
In practice, it is not always feasible to answer such questions when expected outputs need to be identified for a large set of test inputs (e.g., for various resources, roles and privileges). 
Recent incidents involving corporate Web sites, such as Facebook's, indicate that it is particularly difficult to verify, at testing time, large sets of input sequences including the ones that trigger vulnerabilities~\cite{FBviewAs,FBotherVuln}.

Although several security testing approaches have been proposed, 
they typically do not address the oracle problem and assume the availability of an implicit test oracle~\cite{Barr2015}. Furthermore, most approaches focus on a particular vulnerability 
(e.g., buffer overflows~\cite{Haller:2013:DOG,Ognawala-SymbExecutionLowLevelVuln-ASE-2016}) 
and can only uncover vulnerabilities that prevent a system from providing results (e.g., system crashes because of buffer overflows). 

Metamorphic Testing (\MT) is a testing technique which has shown, in some contexts, to be very effective to alleviate the oracle problem~\cite{Chen1998,Liu2014}.
\emph{\MT is based on the idea that it may be simpler to reason about relations between outputs of multiple test executions, called metamorphic relations (\MRs), than it is to specify its input-output behavior}~\cite{Segura2016}. In \MT, system properties are captured as \MRs 
that are used to automatically transform an initial set of test inputs into follow-up test inputs. If the outputs of the system under test for the initial and follow-up test inputs violate the \MR, it is concluded that the system is faulty.

Considerable research has been devoted to developing \MT approaches for application domains such as computer graphics (e.g.,~\cite{Mayer2006,Guderlei2007,Just2009,Kuo2011}), 
Web services (e.g.,~\cite{Chan2007b,Sun2011,Zhou2012}), and embedded systems (e.g.,~\cite{Tse2004,Chan2007,Kuo2011b,Jiang2013}). 
\CHANGED{Unfortunately, only a few approaches target security aspects~\cite{ChenMTSecurity2016}; also, their applicability is limited to the functional testing of security components (e.g., 
code obfuscators~\cite{ChenMTSecurity2016}) or to the verification of specific security bugs (e.g., heartbleed~\cite{Heartbleed}).}
They do not support the specification of general security properties by using \MRs.
Although \MT is automatable, very few \MT approaches provide proper tool support~\cite{Segura2016}. This is also a significant obstacle for tailoring the current approaches for security testing. 
Our goal in this paper is to adopt \MT to address the test oracle problem in security testing. 
Our motivation is to have a systematic way to specify \MRs that capture security properties of Web systems (i.e., properties that are violated only if the system is vulnerable) and to automate security testing by relying on these \MRs. 
\CHANGED{An example of \MR to spot bypass authorization schema vulnerabilities is: 
\emph{a Web system should return different responses to two users when the first user requests a URL that is provided to her by the GUI (e.g., in HTML links) and the second user requests the same URL but this URL is not provided to her by the GUI.}
In other words, a user should not be able to directly access URLs not provided by the GUI.}

In this paper, we propose
an \MT approach that supports engineers in specifying \MRs to capture security properties of Web systems and that automatically detects vulnerabilities (i.e., violations of security properties) based on those relations. 
Our approach is built on top of the following novel contributions:
(1) a Domain-Specific Language (DSL) for specifying \MRs for software security, (2)
a catalog of system-agnostic \MRs targeting well-known security vulnerabilities of Web systems~\cite{OWASPtesting}, (3)
a framework that automatically collects the data required to perform \MT,
and (4) a testing framework that automatically performs security testing based on the \MRs and the collected data. 
To facilitate the specification of \MRs in our DSL, we provide an editor  
which has been implemented as a plug-in for the Eclipse IDE~\cite{EclipseIDE}.

We applied our approach to discover vulnerabilities in a commercial Web system and in Jenkins, a leading open source automation server~\cite{Jenkins}.
The approach automatically detected 100\% and 75\% of the targeted vulnerabilities affecting these two systems, respectively. %
Based on these results and an assessment of the effort involved, we conclude that our approach is practical and beneficial to alleviate the oracle problem in security testing and to automatically detect vulnerabilities in industrial settings. 
Our \MT toolset and the empirical data are publicly available~\cite{WebSMRL}.

This paper is structured as follows. Section~\ref{sec:background} provides the background information regarding \MT. 
Section~\ref{sec:related} discusses the related work. In Section~\ref{sec:approach}, we present an overview of the approach. Sections~\ref{sec:dsl}~to~\ref{sec:mtframework} describe the core technical solutions. Section~\ref{subsec:mrs} presents our catalog of \MRs. 
In Section~\ref{sec:evaluation}, we present the empirical evaluation of our approach. We conclude the paper in Section~\ref{sec:conclusion}.

\section{Background}
\label{sec:background}

In this section, we present the basic concepts of MT.
The core of \MT is a set of \MRs, which are necessary properties of the program under test in relation to multiple inputs and their expected outputs~\cite{Chen2018}. 

In \MT, a single test case run requires multiple executions of the system under test with distinct inputs. 
The test outcome (pass or fail) results from the verification of the outputs of different executions against the \MR. 

As an example, let us consider an algorithm $f$ that computes the shortest path for an undirected graph $G$. For any two nodes $a$ and $b$ in the graph $G$, it may not be practically feasible to generate all possible paths from $a$ to $b$, and then check whether the output path is really the shortest path. However, a property of the shortest path algorithm is that the length of the shortest path will remain unchanged if the nodes $a$ and $b$ are swapped.
Using this property, we can derive an \MR, i.e., $|f(G, a, b)| = |f(G, b, a)|$, in which we need two executions of the function under test, one with $(G, a, b)$ and another one with $(G, b, a)$. The results of the two executions are verified against the relation. If there is a violation of the relation, then $f$ is faulty.

We provide below basic definitions underpinning \MT.

\vspace{0.10cm}
\textit{Definition 1 (Metamorphic Relation - \MR).} Let $f$ be a function under test. A function $f$ typically processes a set of arguments; we use the term \emph{input} to refer to the set of arguments processed by the function under test. 
In our example, one possible input is $(G,a,b)$. The function $f$ produces an output.
An \MR is a condition that holds for any set of inputs $\langle x_{1}, ..., x_{n} \rangle$ where $n \geq 2$, and their corresponding outputs $\langle f(x_{1}), ..., f(x_{n}) \rangle$. 
\MRs are typically expressed as implications.

In our example, the property of the target algorithm $f$ is ``the length of the shortest path will remain the same if the start and end nodes are swapped''. The \MR of this property is 
{\footnotesize{$(x_1=(G, a, b)) \land (x_2=(G, b, a))  \rightarrow  |f(x_1)| = |f(x_2)|$}}.

\vspace{0.10cm}
\textit{Definition 2 (Source Input and Follow-up Input).} 
An \MR implicitly defines how to generate a \emph{follow-up input} from a \emph{source input}. 
A source input is an input in the domain of $f$. A follow-up input is a different input that satisfies the properties expressed by the \MR.
In our example, $(G, a, b)$ and $(G, b, a)$ are the source and follow-up inputs, respectively.

Follow-up inputs can be defined by applying \emph{transformation functions} to the source inputs. The use of \emph{transformation functions} in \MRs simplifies the identification of follow-up inputs.  
In our example, a transformation function that swaps the last two arguments of the source input can be used to define the follow-up input:\\ 
{\footnotesize{$x_1=(G, a, b) \land x_2=\mathit{swapLastArguments}(x_1)  \rightarrow  |f(x_1)| = |f(x_2)|$}} \\
where $\mathit{swapLastArguments}$ is the transformation function.

\vspace{0.10cm}
\textit{Definition 3 (Metamorphic Testing - MT).} MT consists of the following five steps:

\begin{itemize}

\item[1] Generate one source input (or more if required). In our example, a (random) graph $G$ is generated; two vertices $a$ and $b$ in $G$ are randomly selected for the source input. 

\item[2] Derive follow-up inputs based on the \MR. In our example, the function $\mathit{swapLastArguments}$ is applied to $(G, a, b)$.

\item[3] Execute the function under test with the source and follow-up inputs to obtain their respective outputs. In our example, the shortest path function is executed two times with $(G, a, b)$ and $(G, b, a)$.

\item[4] Check whether the results violate the \MR. If the \MR is violated, then the function under test is faulty.

\item[5] Restart from (1), up to a predefined number of iterations.

\end{itemize}

\section{Related Work}
\label{sec:related}

Security testing approaches can be categorized
\cite{Felderer2016} as follows:
(1) security functional testing validating whether the specified security properties are implemented correctly, and 
(2) security vulnerability testing simulating attacks that target typical system vulnerabilities.

Many vulnerability testing approaches rely on an implicit test oracle, i.e., one that relies on implicit knowledge to distinguish between correct and incorrect system behavior~\cite{Haley2008}. 
This is the case for approaches targeting buffer overflows, memory leaks,  unhandled exceptions, and denial of service~\cite{Ognawala-SymbExecutionLowLevelVuln-ASE-2016,Bekrar2011,Takanen2018fuzzing}, \CHANGED{most of which rely on mutational fuzzing~\cite{fuzzingbook2019}, i.e., the generation of new inputs through the random modification of existing inputs.}  
Implicit oracles deal with simple abnormal system behavior such as unexpected system termination and are not system-agnostic. %
What is abnormal in one system might be considered normal in a different context~\cite{Barr2015}.

Vulnerability testing approaches for code injections  also suffer from the oracle problem~\cite{Raghavan2000,Kals2006,Martin2008,Bau2010,Appelt-SLQmutation-ISSTA-2014,Salas2014,Tripp-XSS-ISSTA-2013,Appelt2013}. 
To resolve this problem, Huang et al.~\cite{Huang2003web} proposed an MT-like technique which sends multiple HTTP requests, i.e., one request with an injection, an intentionally invalid request, and a valid request. They compare the responses to determine if the request with the injection is filtered. 
\CHANGED{Unfortunately, MT-like approaches that address a broader set of security vulnerabilities are missing.}

Model-based approaches~\cite{Felderer-MBST-2016,Felderer2011} typically target security vulnerability testing (e.g.,~\cite{Bertolino-PolPa-AST-2012,blome2013vera,he2008attack,Marback2013,jurjens2002umlsec,jurjens2005sound,jurjens2005secure,Jurjens2008,Masood2010,xu2006threat,Martin-PoliciesMutationTestingFramework-WWW-2007,Martin-TestingPoliciesChangeImpact-IWSESS-2007,Martin-PolicyCoverage-ICICS-2006,Wimmel2002,Whittle2008,xu2012automated,xu2015automated}) whereas a few solutions address security functional testing (e.g.,~\cite{Le2007,Mouelhi2008,Mouelhi2009,Xu2012}). 
Most of these approaches only generate test sequences from security models 
and do not address the oracle problem. 
\CHANGED{Approaches that generate test cases including oracles~\cite{xu2012automated,xu2015automated,Xu2012} 
rely on mappings between model-level abstractions (i.e., tokens in markings of PrT networks) and executable code implementing the oracle logic (e.g., searching for error messages in system output). 
Unfortunately, these approaches do not free engineers from implementation effort since they require the manual implementation of the executable oracle code. 
Furthermore, the model-based mapping supported by these approaches does not enable engineers to specify precise test oracles (e.g., oracles that verify the exact content of the output of the system with respect to its inputs~\cite{xu2012automated}).}

\CHANGED{With \MT, we aim to address the limitations of security testing approaches. Indeed, \MT supports oracle automation thanks to \MRs 
that can precisely capture the relations between inputs and outputs.}
Considerable research has been devoted to developing \MT approaches for various domains such as computer graphics (e.g.,~\cite{Mayer2006,Guderlei2007,Just2009,Kuo2011}), simulation (e.g.,~\cite{Chen2009,Ding2011,Murphy2011}), Web services (e.g.,~\cite{Chan2007b,Sun2011,Zhou2012}), embedded systems (e.g.,~\cite{Tse2004,Chan2007,Kuo2011b,Jiang2013}), compilers (e.g.,~\cite{Tao2010,Le2014}), and machine learning (e.g.,~\cite{Xie2009,Murphy2008}). 
\CHANGED{Preliminary applications of MT to security testing~\cite{ChenMTSecurity2016} focus on the functional testing of security components (i.e., verifying the output of code obfuscators and the rendering of login interfaces) and the verification of low level properties broken by specific security bugs (e.g., heartbleed~\cite{Heartbleed}).
Although these works show the feasibility of MT for security, they focus on a narrow set of vulnerabilities and do not automate the generation of executable metamorphic test cases, which are manually implemented based on the identified \MRs.}

 Although \MT is highly automatable, very few approaches provide proper tool support enabling engineers to write system-level \MRs~\cite{Segura2016}. 
\CHANGED{They require that \MRs be defined either as Java methods~\cite{ToolZhu} or pre-/post-conditions~\cite{MurphyJML}, which limit the adoption of \MT to verify system-level, security properties.
Furthermore, since \MRs are often specified by capturing properties using a declarative notation, the use of an imperative language to implement the relations may force engineers to invest additional effort to translate abstract, declarative \MRs.}

\CHANGED{To summarize, existing automated security testing approaches lack support for the generation of test oracles. The few approaches addressing the oracle problem either focus on a limited set of security vulnerabilities, or integrate oracles with limited capabilities.
\MT can overcome these limitations. It can be applied to both security functional testing and vulnerability testing since \MRs can capture both security properties (e.g., a login screen should always be shown after a session timeout) and properties of the inputs and outputs involved in the discovery of a vulnerability (e.g., admin pages are accessed without authentication). 
Existing \MT solutions target few, specific security bugs and do not support automated \MT based on \MRs capturing general security properties. To overcome these limitations, we need a DSL for \MRs and algorithms that automate the execution of \MT.}

\section{Overview of the Approach}
\label{sec:approach}

\begin{figure}[tb]
\hspace{9mm}
\includegraphics[width=7cm]{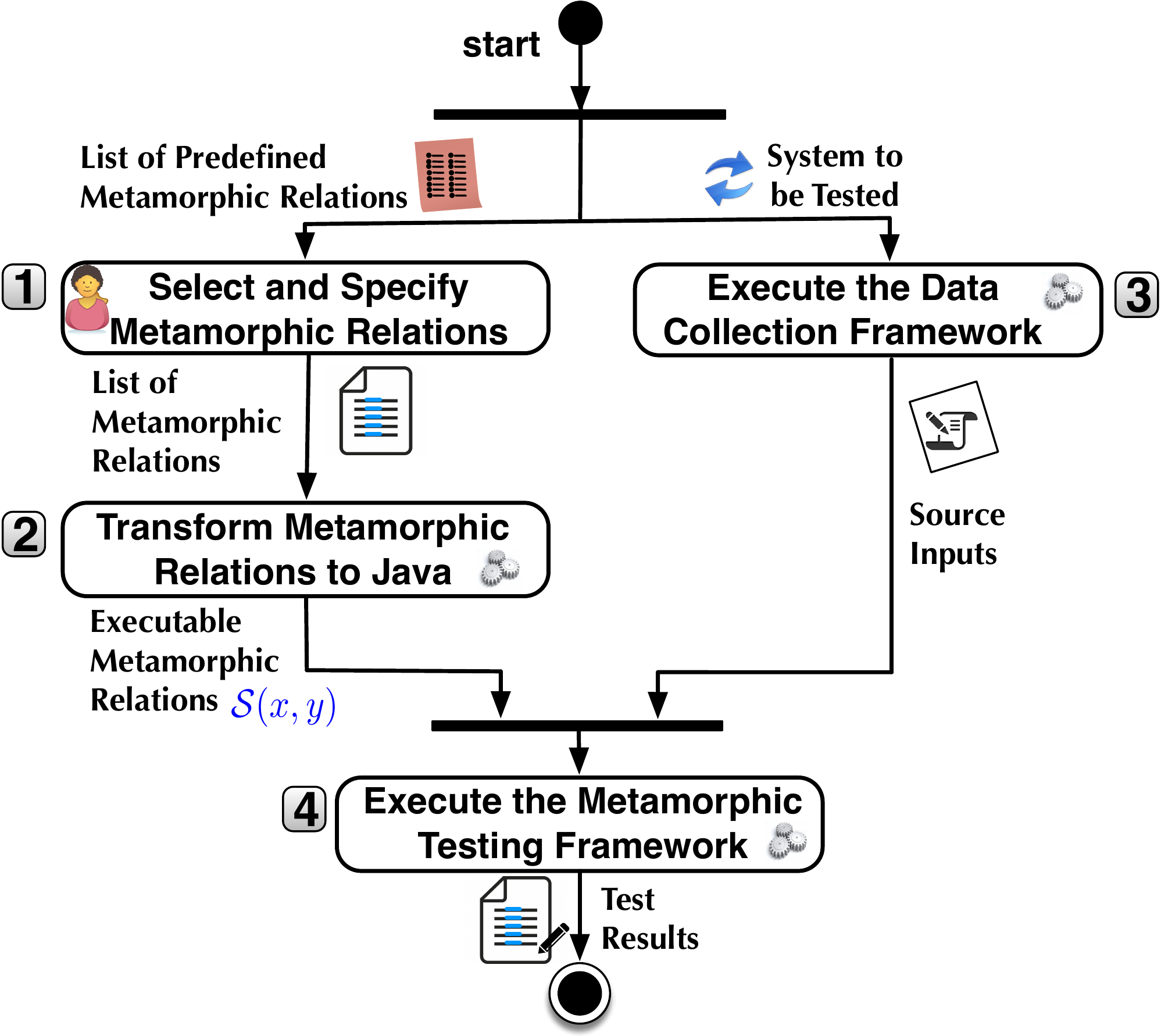}
\caption{Overview of the approach.}
\label{fig:approach}
\end{figure}

The process in Fig.~\ref{fig:approach} presents an overview of our approach. 
In Step 1, the engineer selects, from a catalog of predefined \MRs, the relations for the system under test. 
We have derived our catalog of \MRs from the testing guidelines~\cite{OWASPtesting} edited by OWASP \cite{OWASPWeb}. In addition, the engineer can also specify new relations by using our DSL. Step 1 is manual. We discuss this step in Section~\ref{sec:dsl}.
In Step 2, our approach automatically transforms the \MRs into executable Java code (Section~\ref{sec:transformation}).

In Step 3, the engineer executes a Web crawler to automatically collect information 
about the system under test (e.g., the URLs that can be visited by an anonymous user). 
The crawler determines the structure of the system under test and the actions that trigger the 
generation of new content on a page.
The collected information includes the source inputs for \MT. 
To collect additional information, the engineer can process manually implemented test scripts, if available.
Step 3 does not depend on other steps. 
We discuss Step 3 in Section~\ref{sec:dataframework}.

In Step 4, our approach automatically loads the source inputs required by the \MRs and generates follow-up inputs as described by the relation. After the source and follow-up inputs are executed, their execution results are checked according to the \MRs. The details of the step are described in Section~\ref{sec:mtframework}.

Our DSL and the data collection framework can be extended to support new language constructs and data collection methods.
The \MT framework can be extended to deal with input interfaces not supported yet (e.g., Silverlight plug-ins~\cite{Silverlight}) and to load data collected by new data collection methods.

\section{SMRL: A DSL for Metamorphic Relations}
\label{sec:dsl}

Our approach starts with the activity of selecting and specifying \MRs (Step 1 in Fig.~\ref{fig:approach}). To enable specifying new \MRs, we provide a DSL called Security Metamorphic Relation Language (\SMRL). 
Engineers can also select \MRs for the system under test from the set of predefined \MRs.%

\CHANGED{SMRL is an extension of Xbase~\cite{Efftinge2012xbase}, an expression language provided by Xtext~\cite{Xtext}.}
Xbase specifications can be translated to Java programs and compiled into executable Java bytecode.
We rely on Xbase since DSLs extending Xbase inherit the syntax of a Java-like expression language as well as language infrastructure components, including a parser, 
a linker, a compiler and an interpreter~\cite{Efftinge2012xbase}. These features will facilitate the adoption of \SMRL.

\begin{figure}[tb]
\includegraphics[width=9cm]{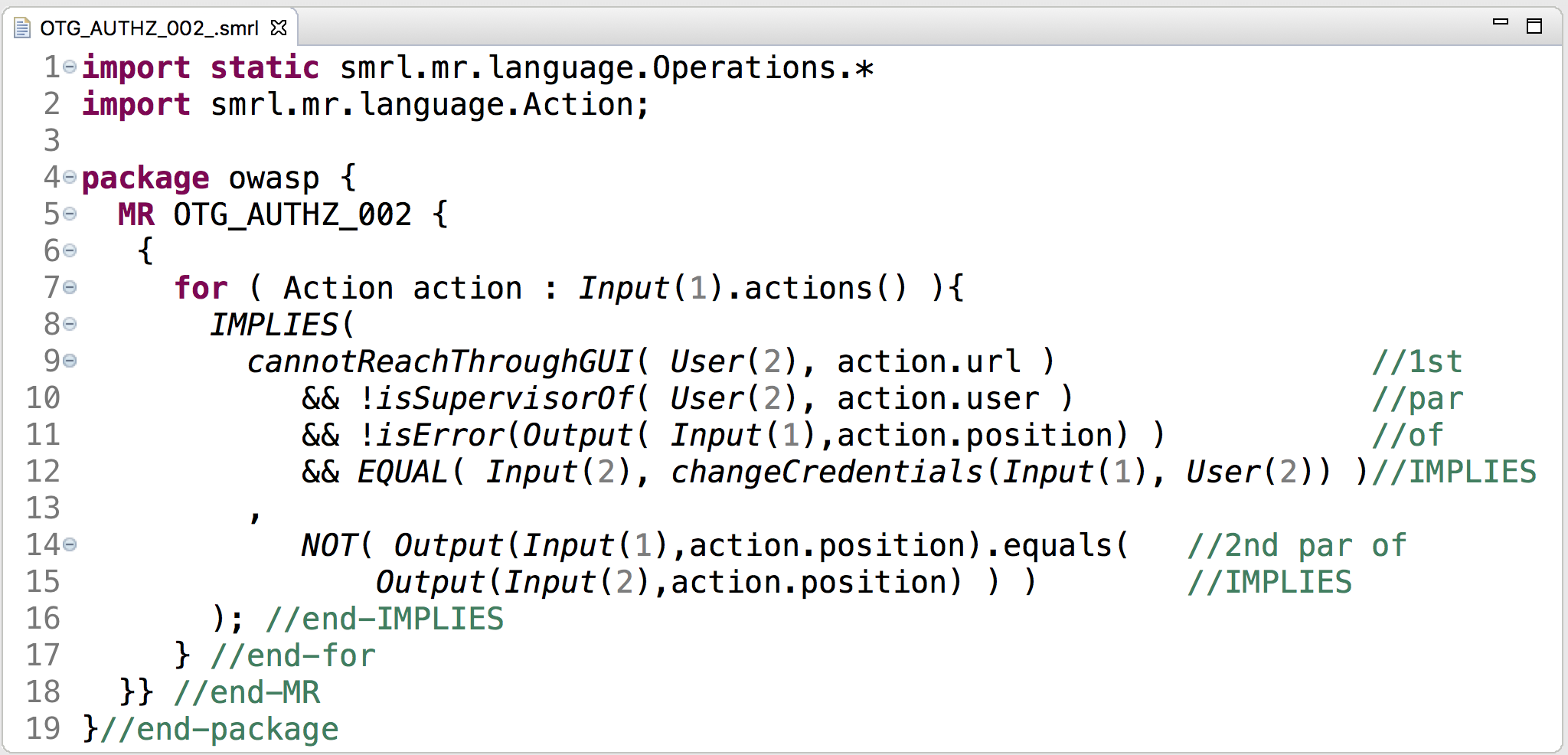}
\vspace*{-2mm}
\captionof{figure}{An \MR for the Bypass Authorization Schema vulnerability.}
\label{fig:mrExample}
\end{figure}

\SMRL extends Xbase by introducing 
(1) a set of data representation functions, %
(2) a set of boolean operators to specify security properties, 
and (3) a set of 
Web-specific 
functions to express data properties and transform data. 
These functions can also be extended by defining new Java APIs to be invoked in \MRs.

Fig.~\ref{fig:mrExample} presents an \MR written in our \SMRL editor. %
The relation checks whether the URLs dedicated to specific users can be accessed by other users through a direct request.
We use it as a running example.

In the following, we introduce the \SMRL grammar, the boolean operators, the data representation functions, and the Web-specific functions.

\subsection{SMRL Grammar}
\label{subsec:grammar}

The SMRL grammar extends the Xbase grammar, which extends the Java grammar.
Each \SMRL specification can have an arbitrary number of import declarations which indicate the APIs to be used in \MRs (Line 1 in Fig.~\ref{fig:mrExample}). 

A package declaration resembles the Java package structure and can contain one or more \MRs.
Line 4 in Fig.~\ref{fig:mrExample} declares the package \emph{owasp}, which is is the package for our \MRs .
Like in Java, \MRs defined in different \SMRL specification files can belong to the same package.

An \MR can contain an arbitrary number of \TEXTTT{XBlock}- \TEXTTT{Expressions}, which are nonterminal symbols defined in the Xbase grammar. An \TEXTTT{XBlockExpression} can contain loops, function calls, operators, and other \TEXTTT{XBlockExpression}s.

\subsection{Data Representation Functions}
\label{subsec:datarep}

SMRL provides 18 functions to represent different types of data 
(i.e., system inputs and outputs) in \MRs. 
Data is typically represented by a keyword followed by an index number used to identify different data items. To keep \SMRL simple, 
we represent data by using functions (hereafter \emph{data functions}) with capitalized names (e.g., \TEXTTT{Input(1)}). 
Table~\ref{table:dataMethods} presents a subset of the data functions in \SMRL. 

\begin{table}[tb]%
\scriptsize
\caption{Excerpt of the data functions in \SMRL.}
\begin{tabular}{|@{\hspace{0.05cm}}p{2.1cm} | @{\hspace{0.05cm}}p{6.2cm} |}
\hline
\textbf{Data function}&\textbf{Description}\\
\hline
Input(int i) & Returns the i\textsuperscript{\emph{th}} input sequence.\\
Action(int i) & Returns the i\textsuperscript{\emph{th}} input action.\\
Session(int i) & Returns the i\textsuperscript{\emph{th}} Web session.\\
User(int i) & Returns the i\textsuperscript{\emph{th}} user of the system.\\
Output(Input i) & Returns the sequence of outputs generated by Input \emph{i}.\\
Output(Input i, int n) & Returns the output generated by the n\textsuperscript{\emph{th}} action of Input \emph{i}.\\
HttpMethod() & Returns the name of an HTTP method (e.g., DELETE).\\
RandomFilePath() & Returns a file system path. We select paths of files in the Web system subfolder, ignoring images, and replacing symbolic links (e.g., `plugins' is mapped to `plugin' in Jenkins).\\
RandomValue(Type~t) & Returns a random value of the given type.\\
\hline
\end{tabular}
\label{table:dataMethods}
\end{table}%

\begin{figure}[tb]
\hspace{5mm}\includegraphics[width=7cm]{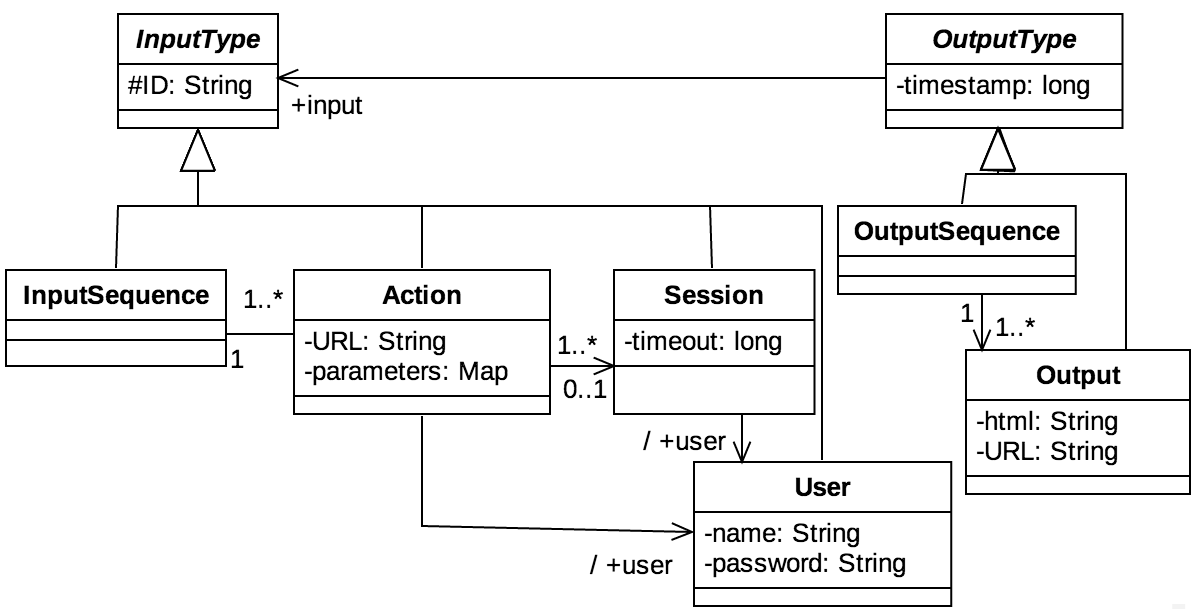}
\caption{Metamorphic data classes in \SMRL.}
\label{fig:dataClasses}
\end{figure}

Each data function returns a data class instance. 
Fig.~\ref{fig:dataClasses} presents the \SMRL data model where all classes are subtypes of either \TEXTTT{InputType} or \TEXTTT{OutputType}.
\TEXTTT{InputType} represents input data that can be defined to trigger a certain system behavior. 
\TEXTTT{InputSequence} represents a sequence of interactions between a user and the system under test and is consequently associated with \TEXTTT{Action}.
\TEXTTT{Action} represents an activity performed by a user %
(e.g., requesting a URL). 
It carries information about actions %
such as a URL requested by an action and parameters %
in the URL query string. 
\TEXTTT{Action} is associated with \TEXTTT{Session}, which represents a user session in a Web application. 
\TEXTTT{User} represents a system user. 

A \emph{source input} is an instance of \TEXTTT{InputType} returned by one of the data functions; 
a \emph{follow-up} input is an instance of \TEXTTT{InputType} modified by means of a Web-specific function (see Section~\ref{subsec:dsl_op}). 
For example, a source input might be a sequence of two HTTP requests for user login and user profile visualization. %
A follow-up input is the same sequence with login credentials for a different user.
Instances of \TEXTTT{OutputType} capture outputs generated by the system when processing an input; each instance of \TEXTTT{OutputType} is associated with an instance of \TEXTTT{InputType}.
The last three functions in Table~\ref{table:dataMethods} return predefined/random values. They are used to redefine attributes of follow-up inputs as described in Section~\ref{sec:mtframework}.
 
\subsection{Boolean Operators}
\label{subsec:boolean}

\SMRL provides seven boolean operators, i.e., \TEXTTT{IMPLIES}, \TEXTTT{AND}, \TEXTTT{OR}, \TEXTTT{TRUE}, \TEXTTT{FALSE}, \TEXTTT{NOT} and \TEXTTT{EQUAL}. 
They enable the definition of \emph{metamorphic expressions}, which are boolean expressions that should hold for an \MR to be true.
A \emph{metamorphic expressions} is a specific kind of \TEXTTT{XBlockExpression}.
We use metamorphic expressions to decompose an \MR into simple properties.
They are defined in a declarative manner, which is standard practice in \MT.

The \MR in Fig.~\ref{fig:mrExample} includes a metamorphic expression using the operator \TEXTTT{IMPLIES}.
Since the expression is within a loop body, 
the relation holds only if the expression evaluates to true in all the iterations over the input actions.

The semantics of the operators \TEXTTT{IMPLIES}, \TEXTTT{AND}, \TEXTTT{OR}, \TEXTTT{TRUE}, \TEXTTT{FALSE}, and \TEXTTT{NOT} is straightforward.
The operator \TEXTTT{EQUAL}, instead, does not simply evaluate the equality of two arguments but defines a follow-up input by assigning the second parameter to the first parameter.
The operator \TEXTTT{EQUAL} acts as an equality operator only when its first parameter refers to an input that has already been used in previous expressions of the \MR. Otherwise, it acts as an assignment operator.
In Fig.~\ref{fig:mrExample}, the operator \TEXTTT{EQUAL} defines the follow-up input \TEXTTT{Input(2)} as a modified copy of \TEXTTT{Input(1)}.

\subsection{Web-Specific Functions}
\label{subsec:dsl_op}

\MRs for security testing often capture complex properties of Web systems that cannot be expressed with simple boolean or arithmetic operators. 
Therefore, \SMRL provides a set of functions that capture typical properties of Web systems and alter Web data. 
Table~\ref{table:dataOperators} describes a portion of the 30 Web-specific functions in \SMRL~\cite{WebSMRL}. %
Each function is provided as a method of the \SMRL API. Engineers can specify additional functions as Java methods. 
The new functions can be used in \SMRL thanks to the underlying Xtext framework.

\begin{table}[tb]
\scriptsize
\caption{Excerpt of the Web-specific functions in \SMRL.}
\begin{tabular}{|@{\hspace{0.1cm}}p{3.6cm} | @{\hspace{0.05cm}}p{4.6cm} |}
\hline
\textbf{Operator}&\textbf{Description}\\
\hline
changeCredentials(Input i, User u) & Creates a copy of the provided input sequence where the credentials of the specified user are used (e.g., within login actions).\\
\hline
copyActionTo(Input i, int from, int to) & Creates a new input sequence where an action is duplicated in the specified position and the remaining actions are shifted by one.\\
\hline
cannotReachThroughGUI( User u, String URL) & Returns true if a URL cannot be reached by the given user by exploring the user interface of the system (e.g., by traversing anchors).\\
\hline
isLogin(Action a) & Returns true if the action performs a login.\\
\hline
isSupervisorOf(User a,User b) & Returns true if `a' can access the URLs of `b'.\\
\hline
afterLogin(Action a) & Returns true if the action follows a login.\\
\hline
isSignup(Action a) & Returns true if the action registers a new user on the system.\\
\hline
isError(Output page) & Returns true if the page contains an error message.\\
\hline
userCanRetrieveContent(User u, Object out) & Returns true if the output data (i.e., the argument `out') has ever been received in response to any of the input sequences executed by the given user during data collection. \\

\hline
\end{tabular}
\label{table:dataOperators}
\end{table}%

The \MR in Fig.~\ref{fig:mrExample} uses the Web-specific functions \TEXTTT{cannotReachThroughGUI}, \TEXTTT{isSupervisorOf}, \TEXTTT{isError} and \TEXTTT{changeCredentials}.
\CHANGED{The relation indicates that 
the same sequence of actions should provide different outputs when performed by two different users under a certain condition. 
The condition is that one of the two users cannot access one of the requested URLs by simply browsing the GUI of the system.
In other words, if the system does not provide a URL to a user through its GUI, then the user should not be allowed to access the URL.
Also, to avoid false alarms, the user who cannot access the URL from the GUI, indicated as \TEXTTT{User(2)} in Fig.~\ref{fig:mrExample}, should not be a supervisor with access to all the resources of the other user, i.e., \TEXTTT{User(1)}.
Finally, we avoid source inputs that return an error message to \TEXTTT{User(1)} because, for these inputs, it is not possible to characterize the output that should be observed for \TEXTTT{User(2)}, who, indeed, may observe the same error, a different error, or an empty page.} 

In Fig.~\ref{fig:mrExample}, the function \TEXTTT{cannotReachThroughGUI} checks if the URL of the current action cannot be reached from the GUI (Line 9).
\CHANGED{The function \TEXTTT{isSupervisorOf}  checks if \TEXTTT{User(2)} is not a supervisor of \TEXTTT{User(1)} (Line 10).
The function \TEXTTT{isError} returns true if an output page contains an error message, based on a configurable regular expressions (Line 11).}
The function \TEXTTT{changeCredentials} creates a copy of a provided input sequence using different credentials. 
It is invoked to define the follow-up input (Line 12).
The data function \TEXTTT{Output} executes the sequence of actions in an input sequence (e.g., requests a sequence of URLs) and returns the output of the i-th action.

\section{SMRL to Java transformation}
\label{sec:transformation}

\SMRL specifications are automatically transformed into Java code (Step 2 in Fig.~\ref{fig:approach}). 
\CHANGED{To this end, we extended the Xbase compiler (hereafter \SMRL compiler).}
Each \MR is transformed into a Java class with the name of the relation and its package.
The generated classes extend the class \TEXTTT{MR} and implement its method \TEXTTT{mr}. %

The method \TEXTTT{mr} executes the metamorphic expressions in the \MR. %
It returns \TEXTTT{true} if the relation holds and \TEXTTT{false} otherwise.
To do so, the \SMRL compiler transforms each boolean operator into a set of nested \TEXTTT{IF} conditions.  %
For example, for the operator \TEXTTT{IMPLIES}, the generated code returns \TEXTTT{false} when the first parameter is true and the second one is false. 
For the case in which the \MR holds, the SMRL compiler generates a statement that returns \TEXTTT{true} at the end of \TEXTTT{mr}.

\begin{figure}[tb]
\includegraphics[width=8.5cm]{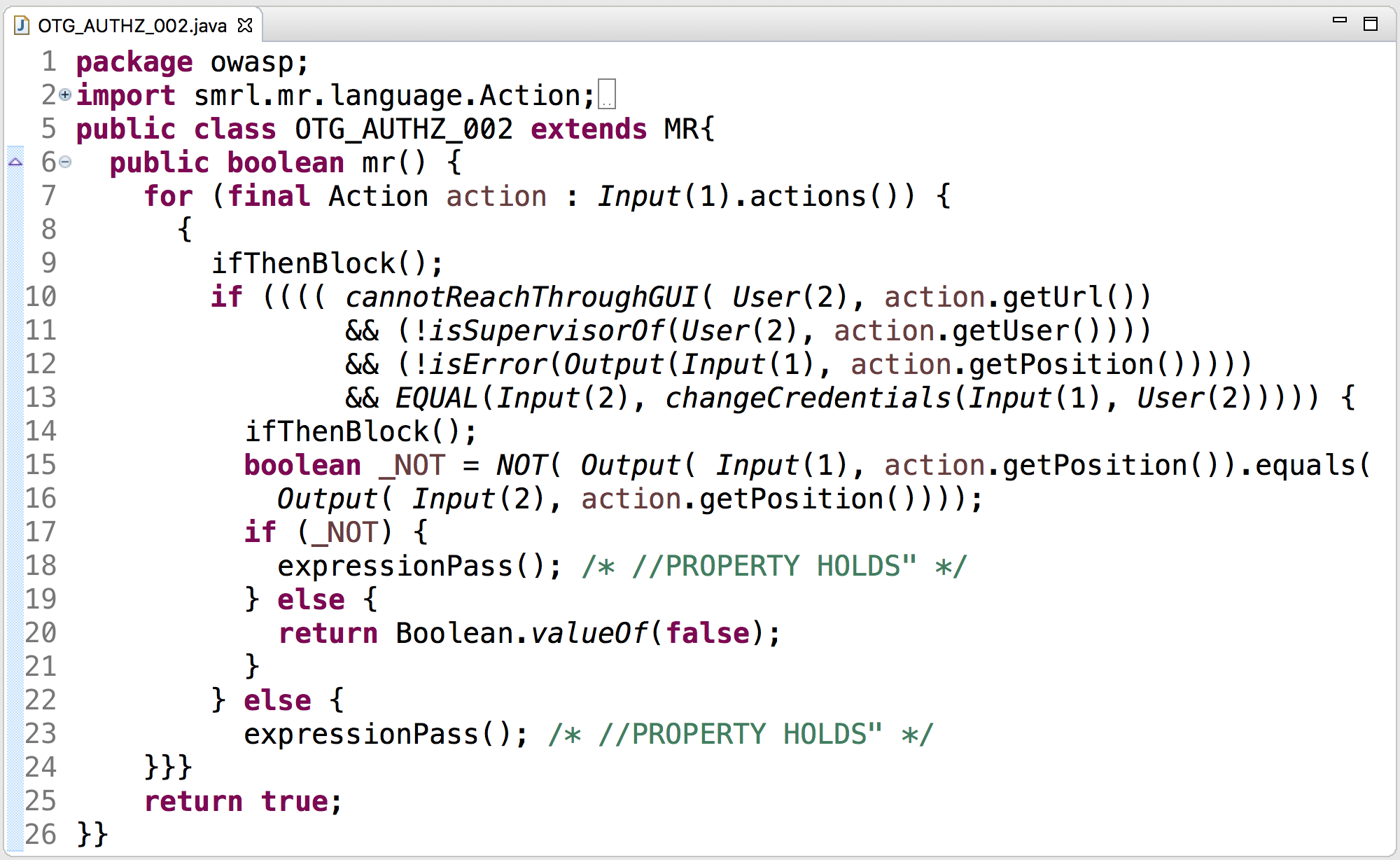}
\caption{Java code generated from the \MR in Fig.~\ref{fig:mrExample}.}
\label{fig:generated002}
\end{figure}

Fig.~\ref{fig:generated002} shows the Java code generated from the relation in Fig.~\ref{fig:mrExample}. 
A loop control structure is generated from the loop instruction in the relation (Line 7). The loop body contains the Java code generated from the metamorphic expression using the operator \TEXTTT{IMPLIES} (Lines 10-24). 
The first \TEXTTT{IF} condition checks whether the first parameter of the operator \TEXTTT{IMPLIES} holds (Lines 10-13). The nested \TEXTTT{IF} block checks whether the second parameter  of \TEXTTT{IMPLIES} holds (Line 17). 
If the expression does not hold, \TEXTTT{mr} returns \TEXTTT{false} (Line~20).
The relation holds only if all the expressions in the loop hold. Therefore, the \SMRL compiler generates a \TEXTTT{return~true} statement after the loop body (Line 25). %
Calls to the methods \TEXTTT{ifThenBlock} and \TEXTTT{expressionPass} are used to erase the generated follow-up inputs at each iteration.

\section{Data Collection Framework}
\label{sec:dataframework}

To automatically derive source inputs (Step 3 in Fig.~\ref{fig:approach}), we extended the Crawljax Web crawler~\cite{Mesbah2012,Mesbah2008}.
 Crawljax explores the user interface of a Web system (e.g., by requesting URLs in HTML anchors or by entering text in HTML forms). It generates a graph whose nodes represent the
 system states reached through the user interface 
 and edges capture the action performed to reach a given state (e.g., clicking on a button). 
Crawljax detects states based on the content of the displayed page. 
\CHANGED{Our extension relies on the edit distance to distinguish system states~\cite{Levenshtein}.} 
We keep a cache of the HTML page associated to each state detected by Crawljax. %
When a new page is loaded, our extension computes the edit distance between the loaded page and all the pages associated to the different system states. When the distance is below a given threshold (5\% of the page length), we assume that two pages belong to the same state. 
If a page does not belong to any state, 
Crawljax adds a new state to the graph.
Crawling stops when no more states are encountered or a timeout is reached.

Our Crawljax extensions enable replicating and modifying portions of a crawling session.
In addition to (i) the Crawljax actions and (ii) the XPath of the elements targeted by the actions (e.g., a button being clicked on), our extension records (iii) the URLs requested by the actions,  (iv) the data in the HTML forms, and (v) the background URL requests.
This enables, for example, replicating modified portions of crawling sessions that request URLs not appearing in the last Web page returned by the system.
To crawl the system under test, we require only 
its URL and a list of credentials.

 \begin{figure}[tb]
\includegraphics[width=8.5cm]{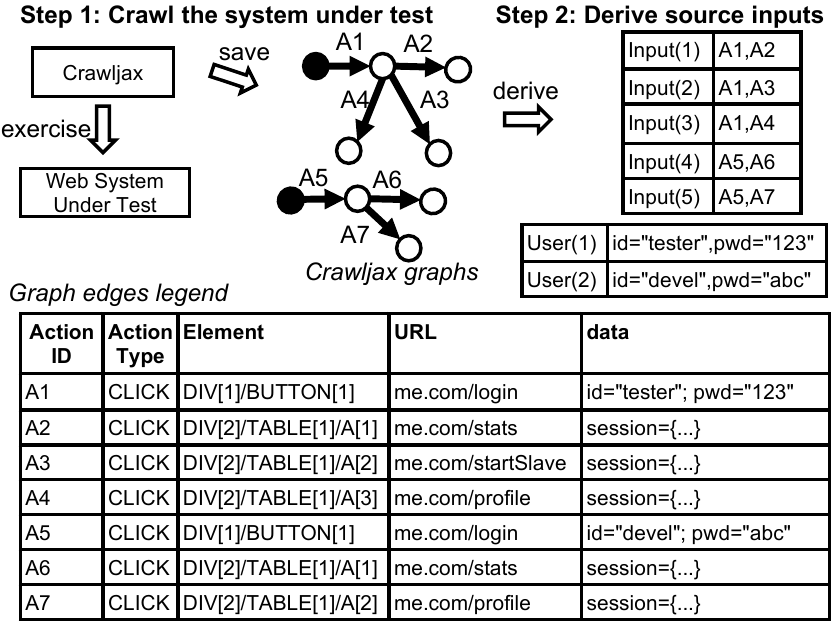}
\caption{Data collection with a simplified example.}
\label{fig:dataCollectionExample}
\end{figure}

\CHANGED{Fig.~\ref{fig:dataCollectionExample} exemplifies the data collection steps. 
First, Crawljax generates the graphs of the system under test.} %
Second, source inputs are automatically derived from the graphs. %
For example, an input sequence is a path from the root to a leaf of a Crawljax graph in depth-first traversal.
The source inputs are later queried by the \SMRL functions (see Section~\ref{sec:mtframework}).
For example, \TEXTTT{Input(i)} returns the $i^{th}$ input sequence; \TEXTTT{User(i)} returns the $i^{th}$ unique login credentials in the input sequences. 

In addition to Crawljax, our toolset also processes manually implemented test scripts to generate additional source inputs.
It processes test scripts based on the Selenium framework~\cite{web:selenium} and derives a source input from each.
We rely on test scripts to exercise complex interaction sequences not triggered by Crawljax (see Section~\ref{sec:evaluation}).
Crawljax, instead, performs an almost exhaustive exploration of the Web interface, which is typically not done by test scripts. 
Engineers can reuse scripts developed for functional testing, or define new ones.

\section{Metamorphic Testing Framework}
\label{sec:mtframework}

We automatically perform testing based on the executable \MRs in Java and the data collected by the data collection framework (Step 4 in Fig.~\ref{fig:approach}). Fig.~\ref{alg:executeMR} presents our testing algorithm. %
The algorithm takes as input a \MR and a data provider exposing the collected data (source inputs).
We first process the bytecode of the \MR to identify the types of source inputs 
referenced by the relation (e.g., \emph{Input} and \emph{User}).
This is achieved by the function \TEXTTT{extractSourceInputTy\-pes} (Line~\ref{alg:executeMR:extract}) which identifies the calls to the \emph{data representation functions} using the ASM static analysis framework~\cite{ASM}.
We ensure that all possible combinations of available source inputs are stressed during the execution of the relation
(e.g., we would like to access all available URLs with all configured users).
This is achieved by the function \TEXTTT{iterate\-Over\-InputTypes} (Line~\ref{alg:executeIterate:OverInputTypes}). The function 
iterates over all available items for a given input type (e.g., all available users) and is recursively invoked for each input type in the \MR.

\begin{figure}[tb]

\begin{algorithmic}[1]

\scriptsize
\Require \emph{MR}, the bytecode of the metamorphic relation to be executed
\Require \emph{dataProvider}, an object that exposes the data collected by the crawlers
\Ensure \emph{Failures}, a list of failing executions with contextual information

\Function{ExecuteMetamorphicTesting}{MR, dataProvider} 
\State srcTypes $\gets$ extractSourceInputTypes(MR) \label{alg:executeMR:extract}
\State iterateOverInputTypes(MR, dataProvider, 0, dataTypes) \label{alg:executeIterate:OverInputTypes}
\State \Return $\mathit{Failures}$
\EndFunction
\Function{iterateOverInputTypes}{MR, dataProvider, i, dataTypes} 
\While {dataProvider.hasMoreViews(dataTypes[i])}  \label{alg:executeMR:hasMore}
\State dataProvider.nextView(dataTypes[i])  \label{alg:executeMR:nextView}
\If{(i $<$ dataTypes.lenght)} \hspace{2mm}\textbf{\textcolor{gray}{//need to iterate over other types}} \label{alg:executeMR:checkAll}
	\State iterateOverInputTypes(MR,dataProvider, i+1,srcTypes)   \label{alg:executeMR:iterateType}
\Else  \hspace{2mm}\textbf{\textcolor{gray}{//we have set a view for every input type in the relation}} \label{alg:executeMR:allViewsSet}
	\State result = MR.run() \textbf{\textcolor{gray}{//execute the metamorphic relation}} \label{alg:executeMR:execute}
	\State \textbf{if} ( result == false)	\textbf{\textcolor{gray}{//the MR does not hold}} \label{alg:executeMR:resultFalse}
	\State  \hspace{5mm}  addFailure(Failures,dataProvider) \textbf{\textcolor{gray}{//trace the failure}} \label{alg:executeMR:addFailure}
\EndIf	
	
\EndWhile
\EndFunction

\end{algorithmic}
\vspace{-3mm}
\caption{Metamorphic testing algorithm.}
\label{alg:executeMR}
\end{figure}

The function \TEXTTT{iterate\-Over\-InputTypes} is driven by the methods exposed by the data provider (Lines~\ref{alg:executeMR:hasMore}~and~\ref{alg:executeMR:nextView}). 
The data provider works as a circular array that provides, in each iteration of \TEXTTT{iterate\-Over\-InputTypes}, 
a different view on the collected data. %
This is achieved through the method \TEXTTT{nextView} (Line~\ref{alg:executeMR:nextView}), which, 
for N input items of a given type (e.g., User),
generates N different views, with items shifted by one position.

After the views are generated, the \MR is executed (Line~\ref{alg:executeMR:execute}). 
Follow-up inputs are generated within the execution of the \MR by the calls to the operator \TEXTTT{EQUAL}.
For example, in Fig.~\ref{fig:mrExample}, the operator \TEXTTT{EQUAL} makes \TEXTTT{Input(2)} refer to a copy of the input sequence returned by the function \TEXTTT{changeCredentials}.

When the relation does not hold (Lines \ref{alg:executeMR:resultFalse} and \ref{alg:executeMR:addFailure}), the function \TEXTTT{addFailure} stores the failure context information (i.e., source-inputs, follow-up inputs, and system outputs). 
To minimize the time spent by engineers in analyzing failures triggered by distinct follow-up inputs exercising a same vulnerability,
we report only failures that perform HTTP requests (e.g., accessing a URL) not generated by input sequences that led to previously reported failures.

Function \TEXTTT{nextView} is iteratively invoked until all the items of a given input type are processed (Line~\ref{alg:executeMR:hasMore}). 
This guarantees that all input item combinations are used. %
For the data functions providing random values, \TEXTTT{nextView} returns 100 different views by default.
Since this may lead to combinatorial explosion, we test each \MR for a maximum of 24 hours.

 \begin{figure}[tb]
\includegraphics{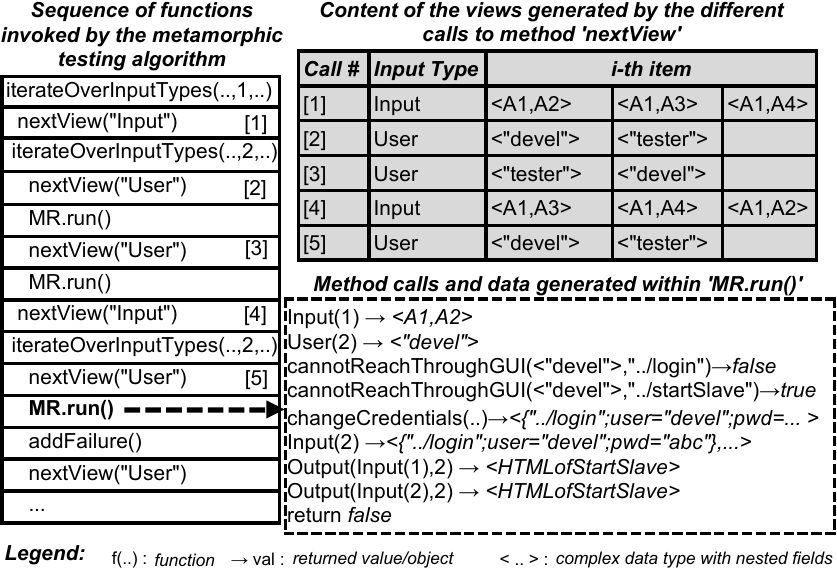}
\caption{Data processing for the relation in Fig.~\ref{fig:mrExample}.} %
\label{fig:simulation}
\end{figure}

\begin{figure}[tb]
\hspace{5mm}\includegraphics[width=8cm]{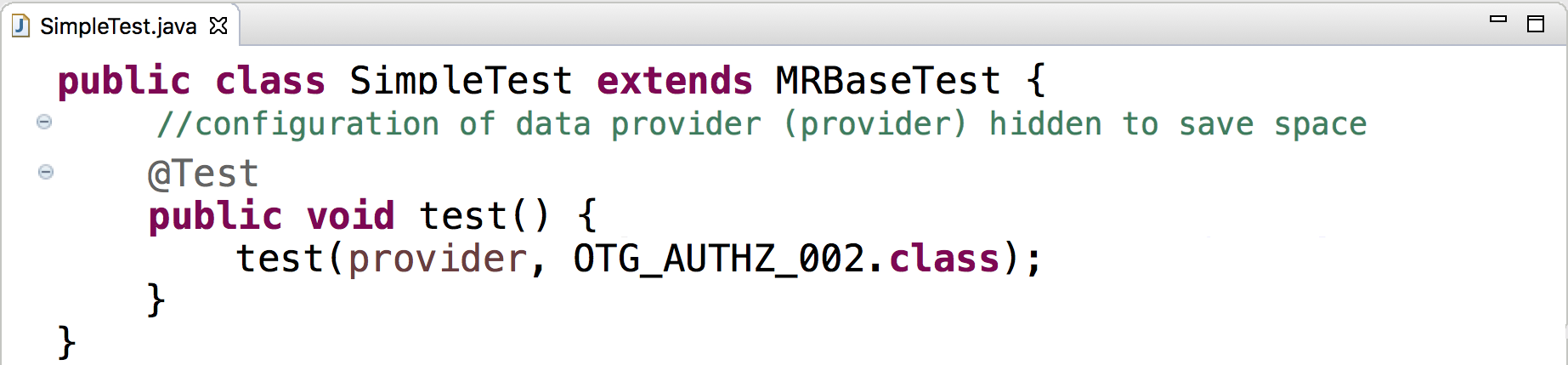}
\caption{Example metamorphic test case. Engineers need only to configure the data provider and select the \MR(s) to be tested.}
\label{fig:exampleTest}
\end{figure}

\CHANGED{Fig.~\ref{fig:simulation} exemplifies the execution of  the relation in Fig.~\ref{fig:mrExample}.}
The table on the left represents the sequence of functions invoked by our algorithm.
In this example, two views for \TEXTTT{User} are inspected for each view of \TEXTTT{Input}.
The first two invocations of \TEXTTT{MR.run} return true (not shown in Fig.~\ref{fig:simulation}) because the \emph{login} and \emph{stats} pages have been accessed by both users \emph{devel} and \emph{tester}
and thus the implication holds. %
The third invocation of \TEXTTT{MR.run} returns false because the output page for the \emph{startSlave} URL is the same for the two input sequences and thus the relation does not hold. To determine if Web pages are equal, 
we rely on edit distance. 
Our framework relies on JUnit~\cite{JUnit} to integrate \MT into traditional testing environments (see Fig.~\ref{fig:exampleTest}).

\section{Catalog of Metamorphic Relations}
\label{subsec:mrs}

We derived a catalog of \MRs %
from the activities described in the OWASP book on security testing~\cite{OWASPtesting}. 
The book provides detailed descriptions of 90 testing activities (hereafter \textit{OWASP testing activities}) for Web systems; each OWASP testing activity targets a specific vulnerability. 
For example, for the bypass authorization schema vulnerability, OWASP suggests to collect links in administrative interfaces and to directly access the corresponding URLs by using credentials of other users. Based on this suggestion, we defined the \MR in Fig.~\ref{fig:mrExample}. 

\begin{table*}[tb]
\scriptsize
\caption{Excerpt of the metamorphic relation catalog for security testing.} 
\begin{tabular}{|@{\hspace{1mm}}p{10.2cm}|@{\hspace{1mm}}p{7.2cm}@{\hspace{0mm}}|}
\hline
\textbf{OTG-AUTHN-001}: Testing for credentials transported over an encrypted channel &
\emph{Description}: A login operation should not succeed if performed on the
\\
\begin{minipage}{10cm}
\vspace{-1mm}
\includegraphics[height=2.75cm]{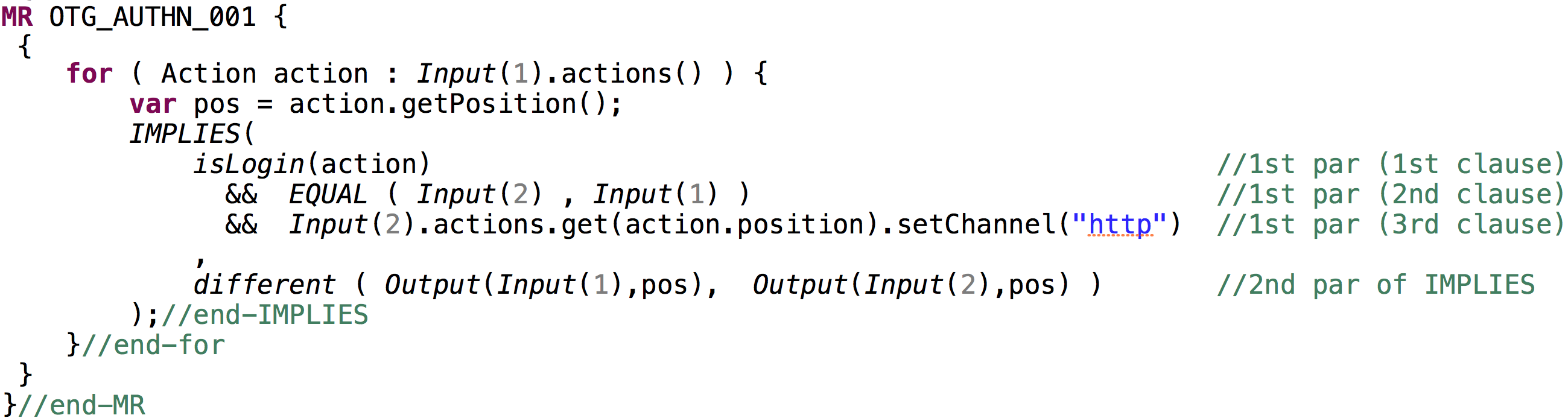} 
\end{minipage}
&
\begin{minipage}{7.2cm}
http channel. The 1st parameter of the operator \texttt{IMPLIES}	is a boolean expression with three clauses joined with logical conjunctions. The 1st clause checks if the current action performs a login. The 2nd clause defines the follow-up input. The 3rd clause changes the channel of the login action in the follow-up input. 
The 2nd parameter of IMPLIES checks if the output generated by the login operation is different in the two cases.
\vspace{9mm}
\end{minipage}
\\
\hline
\textbf{OTG-AUTHZ-001}: Testing for directory traversal/file include&\emph{Description}: A file path passed in a parameter should never enable a 
\\	
\begin{minipage}{10cm}
\includegraphics[height=3.6cm]{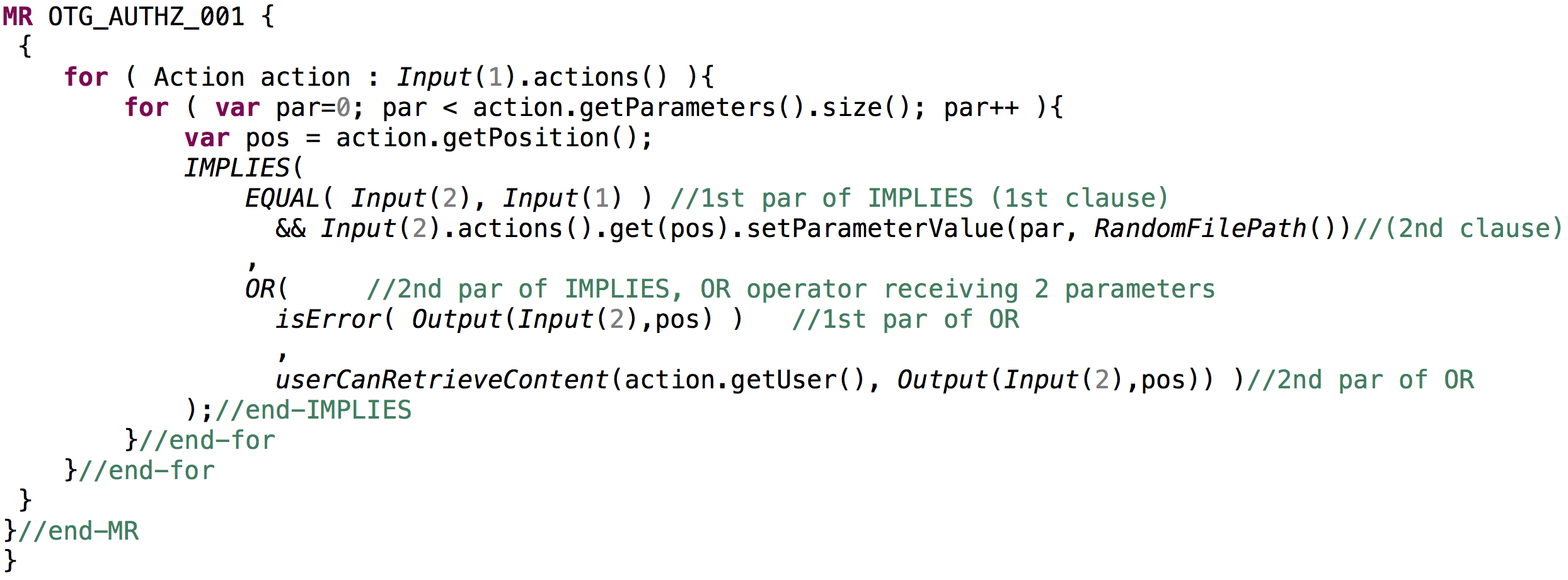}
\end{minipage}
&
\begin{minipage}{7.2cm}
user to access data that is not provided by the user interface.
This metamorphic relation contains two nested loops; the first iterates over the actions in the input sequence, the second iterates over the parameters of the action.
The 1st parameter of the operator \texttt{IMPLIES}	is a boolean expression with two clauses joined with a logical conjunction.
The 1st clause defines a follow-up input that is a copy of the source input.
The 2nd clause set the value of a parameter to a random file path.
The 2nd parameter of IMPLIES verifies the result. It is implemented as an OR operation where the 1st parameter verifies that the follow-up input leads to an error page. 
The 2nd parameter deals with the case in which the generated request is valid, and verifies that the returned content is something that the user has the right to access.
The framework evaluates the \MR as many times as needed to provide 100 different random file paths to the parameters of the action in the position \TEXTTT{pos}.
\end{minipage}	
\\
\hline
\textbf{OTG-SESS-003}: Testing for session fixation&\emph{Description}: A signup action should always lead to a new session ID,\\	
\begin{minipage}{10cm}
\includegraphics[height=3.4cm]{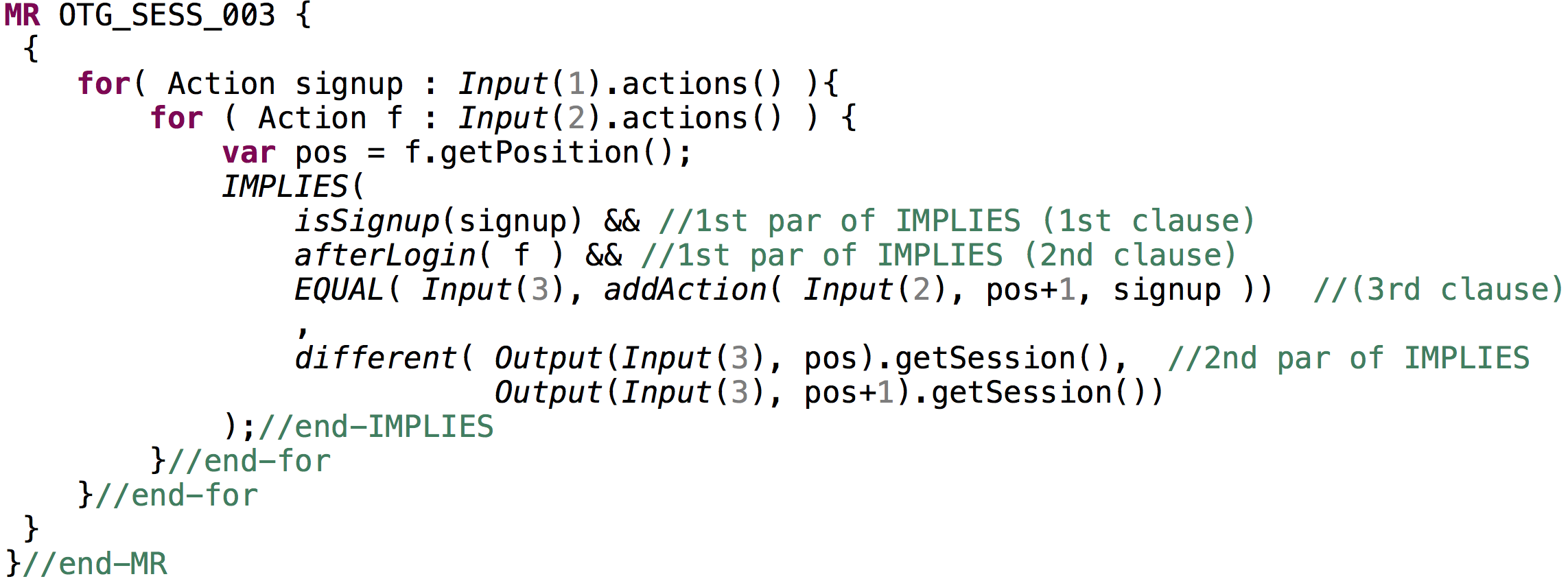}
\end{minipage}
&
\begin{minipage}{7.2cm}
even when performed by a user who is already logged-in.
This metamorphic relation contains two nested loops iterating over the actions of two distinct source input sequences (i.e., \texttt{Input(1)} and \texttt{Input(2)} ). The first loop looks for a signup action (i.e., `signup'), the second looks for an action (i.e., `f') following a login.
The 1st parameter of the operator \texttt{IMPLIES}	is a boolean expression with three clauses joined with a logical conjunction.
The 1st clause checks if we are in the presence of a signup action.
The 2nd clause checks if the action `f' follows a login. 
The 3rd clause defines a follow-up input by copying the signup action after the action `f' in the source input \texttt{Input(2)}.
The 2nd parameter of IMPLIES verifies the result by checking that the session ID following the signup action is different than the one of the previous page.
\vspace{1mm}
\end{minipage}	
\\
\hline
\multicolumn{2}{l}{\hspace{-2mm}\begin{minipage}{17.7cm}
\vspace{2mm}
\textbf{Notes:} Our catalog of metamorphic relations covers also the following OWASP activities: 
testing for HTTP Strict Transport Security (OTG-CONFIG-007),
testing for weaker authentication in alternative channel (OTG-AUTHN-010),
testing for privilege escalation (OTG-AUTHZ-003),
testing for bypassing authentication schema (OTG-AUTHN-004), 
testing for insecure direct object references (OTG-AUTHZ-004),
testing for logout functionality (OTG-SESS-006),
test session timeout (OTG-SESS-007),
testing for Session puzzling (OTG-SESS-008),
testing for HTTP verb tampering (OTG-INPVAL-003),
testing for HTTP parameter pollution (OTG-INPVAL-004),
testing for weak encryption (OTG-CRYPST-004),
test number of times a function can be used (OTG-BUSLOGIC-005), 
test for bypass authorization schema (OTG-AUTHZ-002, see Fig.~\ref{fig:mrExample}). 
\end{minipage}}

\end{tabular}
\label{table:MRcatalog}
\end{table*}

Some OWASP testing activities can be performed in multiple ways. Therefore, we have multiple relations for those activities. Also,
not all the OWASP testing activities benefit from \MT. \CHANGED{The capabilities of \MT are discussed in Section~\ref{sec:evaluation}.}
We defined 22 \MRs which automate 16 OWASP activities. 

\CHANGED{The MRs in our catalog rely on the observation that security testing might be performed using follow-up inputs that cannot be generated by interacting with the GUI of the system but conform with the input format of the system and match its configuration (e.g., the URLs requested by the unauthorized user refer to existing system resources).
We inherit from mutational fuzzing the idea of generating follow-up inputs by altering valid source inputs. However, to generate inputs that are both valid and match the system configuration, instead of relying on random values, we alter source inputs using the data provided by the SMRL Web-specific functions, which return domain-specific information (e.g., protocol names) and crawled data.
Finally, by capturing properties of the output generated by source and follow-up inputs we identify vulnerabilities that cannot be detected with implicit oracles.} 

Table~\ref{table:MRcatalog} presents an excerpt of our catalog along with a description of each \MR. 
The full catalog of \MRs is available for download~\cite{WebSMRL}.
All the \MRs in the catalog are expressed by means of an implication (the operator \TEXTTT{IMPLIES}). %
The operator \TEXTTT{EQUAL} is used to define follow-up inputs.
It indicates that the follow-up input (typically \TEXTTT{Input(2)}) is a copy of the source input (usually \TEXTTT{Input(1)}) except for the differences made by the function calls following the operator.
For example, in \TEXTTT{OTG\_AUTHN\_001}, the follow-up input is equal to the source input except for one action of the input sequence which should be performed on the HTTP channel.

All the \MRs include a loop, which enables defining multiple follow-up inputs
by iteratively modifying different actions of the source input. 
For example, \TEXTTT{OTG\_AUTHN\_001} works with all the login actions observed in the source input sequence. %
The function \TEXTTT{isLogin()} returns true only if the current action performs a login; 
\CHANGED{otherwise, the implication trivially holds and no follow-up input is generated.}

In our catalog, the right-hand side of the implication usually captures the relation between the outputs of the source and follow-up inputs. In \TEXTTT{OTG\_AUTHN\_001}, it is implied that the output for the follow-up input (which performs a login on the unencrypted HTTP channel) should be different than the output for the source input because it should not be possible to login using the HTTP channel.

\section{Evaluation}
\label{sec:evaluation}

Our evaluation addresses the following research questions:

\textbf{\emph{RQ1. To what extent can metamorphic testing address the oracle problem in the context of security testing?}}
We aim to determine which types of security vulnerabilities can be addressed by our solution.

\textbf{\emph{RQ2. Is the proposed solution effective?}} 
The goal is to assess whether the proposed solution enables, in a reliable manner, the automated detection of security vulnerabilities.

\vspace{1mm}
\noindent \textbf{RQ1} 
To answer RQ1, we analyzed the security testing activities recommended by OWASP~\cite{OWASPtesting}. For each activity, we identified state-of-the-art oracle automation strategies. %
Table~\ref{table:OWASPtestingActivities} lists the number of activities automated by these strategies.

\emph{Implicit oracle.} Some activities can be automated by random test input generation strategies relying on implicit oracles. 
For instance, \emph{testing for buffer overflow} 
\cite{OTG:INPVAL:014}
is automated by looking for system crashes in response to lengthy inputs. 

\emph{Catalog-based.} We can automate some activities based on a predefined catalog in which we specify inputs and oracles. 
For instance, we can use a catalog to perform a dictionary attack
for \emph{testing for default credentials}~\cite{OTG:AUTHN:002}.

\begin{table}[tb]
\scriptsize
\begin{minipage}[t]{4.35cm}
\caption{Oracle automation strategies for security testing*}
\begin{tabular}{|@{\hspace{0.05cm}}p{2.4cm} | @{\hspace{0.05cm}}p{1.3cm}|}
\hline
\textbf{Oracle automation strategy}&\textbf{\# OWASP activities automated}\\
\hline
Implicit oracle	& 2\\
Catalog-based	& 6\\
No oracle needed	 &19\\
Manual oracle	& 25\\
Vulnerability-specific & 22\\
Metamorphic testing & 16\\
\hline
\end{tabular}
\label{table:OWASPtestingActivities}
\vspace{0.1cm}
\footnotesize{*Details are available online~\cite{WebSMRL}.}
\end{minipage}
\begin{minipage}[t]{4.4cm}
\caption{Vulnerability types addressed by SMRL \MRs*}
\begin{tabular}{|@{\hspace{0.05cm}}p{3.3cm}|@{\hspace{0.05cm}}p{6mm}|}
\hline
\textbf{Vulnerability type}			&\textbf{\#MRs}\\
\hline
Injection				&0\\
Broken Authentication	&6\\
Sensitive Data Exposure	&5\\
XML External Entities(XEE)&0\\
Broken Access Control	&7\\
Security Misconfiguration	&3\\
Cross-site scripting	(XSS)	&0\\
Insecure Deserialization	&0\\
Vulnerable Components	&1\\
Insufficient Logging	&0\\
\hline
\end{tabular}
\label{table:OWASP10}
\end{minipage}
\end{table}%

\emph{No oracle needed.} 
Some activities collect data to reverse engineer the system under test. 
They do not verify security properties of the system and thus do not have an oracle problem.
For instance, the activity \emph{mapping application architecture}~\cite{OTG:INFO:010} identifies the components of a Web system.

\emph{Manual oracle.} Some activities require humans to determine vulnerabilities based on system specifications.
For instance, when \emph{testing for the circumvention of work flows}~\cite{OTG:BUSLOGIC:006} on pay-per-view systems, 
only a human can decide if pending transactions should grant service access, based on specifications.

\emph{Vulnerability-specific approaches.} %
Some activities can be automated by state-of-the-art tools such as Burp Suite (BS)~\cite{BURP} and thus may not necessarily benefit from \MT.
These are the OWASP testing activities that detect cross site scripting and code injection vulnerabilities. %
Other activities are either not targeted or partially automated.
\CHANGED{For example, BS does not automate oracles for OTG-AUTHZ-002~\cite{BurpSuiteAC}. BS enables engineers to compare the content of site maps~\cite{BurpSuiteScanner} recorded in different user sessions (e.g., with and without certain privileges). Unfortunately, it requires that engineers manually identify the privileged resources and inspect the differences in the observed system outputs, which is error prone (e.g., overlooking privileged resources) and expensive.
Even BS plug-ins using Crawljax to build site maps do not address the oracle problem but generate JUnit tests that simply retrieve the mapped resources~\cite{BurpSuiteCSJ}.
With SMRL, engineers, instead, can focus on the specification of system-level properties without performing manual testing activities. Testing activities, including oracles, are automated by the \MT framework.}

\emph{Metamorphic testing.} 
All the other OWASP testing activities not addressed by the approaches above
can be automated by \MT. 
In general, these activities verify if a resource of the system under test can be accessed under circumstances that should prevent it (e.g., unauthenticated user or unencrypted channel).
They benefit from \MT 
since such activities entail the verification of all system resources, which are numerous 
and present specific security properties (e.g., each Web page might be accessed by a different set of users). %
For these activities, we provide a set of \MRs (see Table~\ref{table:MRcatalog}). %

Based on our analysis, out of 90 OWASP testing activities, 19 are not affected by the oracle problem, 30 are automated by state-of-the-art approaches, and 41 cannot be addressed by existing approaches. \MT can automate 16 (39\%) of these 41 activities. Therefore, \textit{we conclude that \MT can play a key role in addressing the oracle problem in security testing}.

To further characterize our catalog of \MRs, we report in Table~\ref{table:OWASP10} the number of MRs targeting the vulnerability types in the OWASP top ten list~\cite{OWASPten}. 
The \MRs in our catalog can discover five of these ten vulnerability types, and thus \textit{have a broad applicability scope}. Note that MRs can discover injection vulnerabilities~\cite{Huang2003web} and, potentially, also XSS and XEE because they all concern injected code. 
In this paper we specifically target vulnerabilities not addressed by existing oracle automation approaches, which is the reason why we ignored injections. 
We leave the investigation of other vulnerability types to future work.

\newcommand{\TENT}[1]{\textcolor{black}{#1}}

\newcommand{\IWS}{E2\xspace}

\newcommand{\JVers}{2\xspace}
\newcommand{\JVuln}{20\xspace}
\newcommand{\JVulnUsed}{8\xspace}
\newcommand{\TotalCred}{\TENT{4}\xspace}
\newcommand{\JCred}{\TENT{four}\xspace}
\newcommand{\IWSCred}{\TENT{two}\xspace}

\newcommand{\JInputSequences}{\TENT{\CHANGEDLAST{156}}\xspace}
\newcommand{\IWSInputSequences}{\TENT{73}\xspace}

\newcommand{\CrawlTimeMax}{\TENT{\CHANGEDLAST{300}}\xspace}
\newcommand{\CrawlTimeJenkins}{\TENT{\CHANGEDLAST{1000}}\xspace}
\newcommand{\CrawlTimeIWS}{\TENT{40}\xspace}

\newcommand{\JDISC}{\TENT{10}\xspace}

\vspace{1mm}
\noindent \textbf{RQ2} 
We applied the proposed approach to discover vulnerabilities in two case studies: a commercial Web system developed in the context of the EDLAH2 project~\cite{EDLAH} (hereafter, \IWS) and Jenkins~\cite{Jenkins}, an open source system. 
\IWS is the entry point of a healthcare service developed by our industry partner~\cite{MiCare}.
It relies on mobile and wearable devices to support elderly people (patients) in their daily life. 
The \IWS Web interface enables carers (e.g., family and doctors) to monitor patients' conditions. 
The second case study, Jenkins, is an open-source continuous integration server. 
We chose Jenkins since it is  widely adopted and well tested. Its Web interface includes advanced features such as Javascript-based login and AJAX interfaces. 
The two case studies are therefore very different and provide complementary perspectives.
\IWS is developed in PHP~\cite{PHP} and based on the Drupal content management system~\cite{Drupal}; Jenkins is a Java Web application that can be executed within any servlet container~\cite{Jetty}.
We used the latest \IWS version and Jenkins version 2.121.1. 
We selected the Jenkins version affected by all the vulnerabilities triggerable from the Web interface, discovered in 2018, and reported in the CVE vulnerability database~\cite{CVE} after June 1st, 2018. 
Jenkins 2.121.1 is affected by 20 such vulnerabilities.
\IWS is affected by 12 vulnerabilities discovered by manual testing following the OWASP guidelines~\cite{ISSRE}.

Our approach addresses \CHANGEDLAST{36}\% (\CHANGEDLAST{4} out of 11) and 40\% (\JVulnUsed out of \JVuln) of the vulnerabilities affecting \IWS and 
Jenkins, respectively. This is consistent with our analysis in RQ1.

For each system under test, we configured our data collection framework with multiple users having different roles. We used \IWSCred credentials for \IWS and \JCred credentials for Jenkins. For each role, we executed the data collection framework to crawl the system under test for a maximum of \CrawlTimeMax minutes. In total, the data collection took \CrawlTimeJenkins minutes for Jenkins and \CrawlTimeIWS minutes for \IWS. 
\CR{For \IWS and for the anonymous role in Jenkins, Crawljax completed in less than \CrawlTimeMax minutes because all states were visited.}
\IWSInputSequences and \JInputSequences input sequences were identified for \IWS and Jenkins, respectively.
Also, we implemented Selenium-based test scripts to exercise use cases not covered by Crawljax. 
This led to one and two test scripts for \IWS and Jenkins, respectively.
We tested the two systems against the \MRs that target the vulnerabilities affecting them (4 for \IWS and 8 for Jenkins).
Our replicability package~\cite{WebSMRL} does not include \IWS data because of confidentiality restrictions.
\CHANGED{Comparing with state-of-the-art tools is infeasible because they do not provide automated oracles.}

\begin{table}[tb]
\scriptsize
\caption{Summary of RQ2 results grouped by data collection method.}
\begin{tabular}{|@{\hspace{0.1cm}}p{1.6cm} @{\hspace{0.05cm}}| @{\hspace{0.05cm}}p{1.5cm} |  @{\hspace{0.08cm}}r@{\hspace{0.08cm}} | @{\hspace{0.08cm}}r@{\hspace{0.08cm}}    |  @{\hspace{0.08cm}}r@{\hspace{0.08cm}} | @{\hspace{0.08cm}}r@{\hspace{0.08cm}} |  }
\hline
\multirow{2}{*}{\textbf{Case study}}&\multirow{2}{*}{\textbf{Vulnerabilities}}&\multicolumn{2}{c}{\textbf{Crawljax}}&\multicolumn{2}{c|}{\textbf{Crawljax \& Manual}}\\
&&\textbf{Specificity}&\textbf{Sensitivity}&\textbf{Specificity}&\textbf{Sensitivity}\\
\hline
\IWS	& \CHANGEDLAST{4} \cite{ISSRE}& 100.00\% & \CHANGEDLAST{75.00}\% & 100.00\% & 100.00\%\\
\hline
Jenkins 2.121.1	& \JVulnUsed \cite{JV1,JV2,JV3,JV4,JV5,JV6,JV8,JV9}& \CHANGEDLAST{99.34}\% & 50.00\% &   \CHANGEDLAST{99.43}\% & 75.00\%  \\
\hline
\textbf{Overall} & \CHANGEDLAST{12} & \CHANGEDLAST{99.43}\% & \CHANGEDLAST{58.33}\% &   \CHANGEDLAST{99.50}\% & \CHANGEDLAST{83.33}\%  \\
\hline
\end{tabular}
\label{table:resultsRQ2}
\end{table}%

We measured specificity and sensitivity~\cite{Lane2003}. 
Specificity (i.e., the true negative rate) is the ratio of follow-up inputs, generated by our framework, that do not trigger any vulnerability and (correctly) do not lead to any MT failure. In other words, \emph{1 - specificity} measures the time spent by engineers on unwarranted MT failures.
Sensitivity (i.e., the true positive rate) is the ratio of vulnerabilities being discovered. 
Based on the existing vulnerability reports for the two systems considered, we
identified the inputs that should uncover vulnerabilities.  MT failures are expected for these inputs to be true positives.
For each MT failure, we manually verified if the test input actually triggered any vulnerability (true positive).
Table~\ref{table:resultsRQ2} summarizes the results obtained with different data collection methods (i.e., based on Crawljax only or integrating Crawljax and manual test scripts).
Each \MR was tested in less than 12 hours, except one which was stopped after 24 hours.
Performance optimizations are part of our future work.

We observe that the approach has \textbf{extremely high specificity} (\CHANGEDLAST{99.50}\%), which indicates that 
only a negligible fraction of follow-up inputs inspected lead to false alarms (32 out of 6401, {\hbox{$\scriptstyle\mathtt{\sim}$}}\CHANGEDLAST{0.5}\%).
False alarms are due to limitations in Crawljax, which, in the case of Jenkins, did not traverse all the URLs provided by the GUI, for all the users. 
Consequently, \MRs concerning authorization vulnerabilities fail.
However, 
it is easy to determine that the URLs causing the false alarms should be accessible to all the users.

\textbf{Sensitivity is high} when data collection is based on both Crawljax and manual test scripts (100\% for \IWS and 75\% for Jenkins).
Since sensitivity reflects the fault detection rate (i.e.,  the portion of vulnerabilities discovered), we conclude that our approach is \textbf{highly effective}. 
Overall, it detects \CHANGEDLAST{83.33}\% of the vulnerabilities targeted in our evaluation.
More precisely, the approach identifies 47 distinct inputs sequences triggering these vulnerabilities.
The approach misses two of the eight targeted vulnerabilities in Jenkins. One of them can be detected only if the server configuration is modified during test execution~\cite{JV3}, which is not supported by our toolset.
The other one cannot be reproduced since it concerns the termination of Jenkins' reboot~\cite{JV1}, which is not interruptible when Jenkins is not overloaded 
(our case).

When the data collection relies on Crawljax only, sensitivity drops below 75\% for Jenkins. 
This occurs since Jenkins requires quick system interactions 
to exercise certain features (e.g., first writing a valid Unix command in a textbox to enqueue a batch job, and then quickly pressing a button to delete it from the queue).  
However, even when the data collection is based on Crawljax only, the overall fault detection rate is satisfactory (i.e., \CHANGEDLAST{58.33}\%), with \CHANGEDLAST{7} out of \CHANGEDLAST{12} vulnerabilities being detected.  
Automatically detecting \CHANGEDLAST{58.33}\% of the vulnerabilities not targeted by state-of-the-art approaches, without the need for any manual test script, is encouraging.

The benefits of our approach mostly stems from the \MRs in our catalog being reusable to test any Web system.
Furthermore, the required manual test scripts are few and inexpensive to implement.
For the Web systems above, we manually wrote three test scripts which only include 10 actions in total.
This is very limited in comparison to the  total of \CHANGEDLAST{6401} inputs sequences (\CHANGEDLAST{41834} actions) automatically generated by our approach  to test the two systems. 
A traditional way to verify the same scenarios would require \CHANGEDLAST{6401} manually implemented test scripts, each providing a distinct input sequence, and a dedicated oracle (e.g., an assertion statement).
Therefore, we conclude that our approach provides an advantageous cost-effectiveness trade-off compared to current practice.

\noindent \textbf{Threats to Validity} The main threat to validity in our evaluation concerns the generalizability of the conclusions.
Regarding RQ1, to mitigate this threat and minimize the risk of considering a set of testing activities that is not representative of Web application testing, we considered testing activities proposed by a third party organization (i.e., OWASP).
As for RQ2, to mitigate this threat, we selected systems that are
representative of modern Web systems but are very different from both a technical and process perspective.

\section{Conclusion}
\label{sec:conclusion}

In this paper, we presented an approach that enables engineers to specify metamorphic relations (\MR) capturing security properties of Web systems, and that automatically detects security vulnerabilities based on those relations. Our approach aims to alleviate the oracle problem in security testing. 

Our contributions include (1) a DSL and supporting tools for specifying \MRs for security testing, (2) a set of \MRs inspired by OWASP guidelines, (3) a data collection framework crawling the system under test to automatically derive input data, %
and (4) a testing framework automatically performing security testing based on the \MRs and the input data~\cite{WebSMRL}.

Our analysis of the OWASP guidelines shows that our approach can automate 39\% of the security testing activities not currently targeted by state-of-the-art techniques, which indicates that the approach significantly contributes to addressing the oracle problem in security testing.
Our empirical results with two commercial and open source case studies show that the approach requires limited manual effort and detects 83\% of the targeted vulnerabilities, thus suggesting it is highly effective.

\vspace{5mm}
\footnotesize{
\section*{Acknowledgment} This work has received funding from the National Research Fund, FNR, Luxembourg, with grant INTER/AAL/15/11213850, from the European Research Council (ERC) under the European Union's Horizon 2020 research and innovation programme (grant agreement No 694277), and from the Canada Research Chair programme.
}


\bibliographystyle{IEEEtran}
\bibliography{biblio}

\end{document}